\documentclass[12pt]{article}%
\usepackage{amsmath}
\usepackage{eurosym}
\usepackage{amssymb}
\usepackage{amsfonts}
\usepackage[onehalfspacing]{setspace}
\usepackage{ulem}
\usepackage{xcolor}
\usepackage{graphicx}%
\providecommand{\U}[1]{\protect\rule{.1in}{.1in}}

\newtheorem{proposition}{Proposition}
\newtheorem{remark}{Remark}

\begin{document}

\title{Monopolistic Data Dumping\thanks{Financial support from the Foerder Institute
and UKRI Frontier Research Grant no. EP/Y033361/1 is gratefully acknowledged.
We thank Tova Milo for helpful comments. We also thank Alex Clyde, Yahel
Menea, Emiliano Sandri, and Chet Geppetti for excellent research assistance.
Finally, we acknowledge error-correction assistance from refine.ink.}%
\linebreak}
\author{Kfir Eliaz and Ran Spiegler\thanks{Eliaz: Tel Aviv University and King's
College London. Spiegler: Tel Aviv University and University College London}}
\maketitle

\begin{abstract}
A profit-maximizing monopolist curates a database for users seeking to learn a
parameter. There are two user types: \textquotedblleft
Nowcasters\textquotedblright\ wish to learn the parameter's current value,
while \textquotedblleft forecasters\textquotedblright\ target its long-run
value. Data storage involves a constant marginal cost. The monopolist designs
a menu of contracts described by fees and data-access levels. The
profit-maximizing menu offers full access to historical data, while current
data is fully provided to nowcasters but may be withheld from forecasters.
Compared to the social optimum, the monopolist keeps too much historical data,
too little current data, and may store too much data overall.\bigskip
\bigskip\bigskip\bigskip

\end{abstract}

\section{Introduction}

Production of digital data has exploded in recent years, primarily because
data usage has expanded to include consumption of both textual and
audio-visual content, individual-specific information that facilitates
targeted advertising, training predictive AI models, etc. Indeed, the current
pace of data production may overtake our capacity to \textit{store} it.
According to a recent report by the IT firm Lightedge,\smallskip

\begin{quote}
\textquotedblleft The International Data Corporation (IDC) predicts that the
world's data will surpass 175 zettabytes by 2025 -- more than tripling the
volume of stored data in 2020. That's a lot of data -- 1 zettabyte alone would
consume enough data center space to fill roughly 20\% of
Manhattan.\textquotedblright%
\footnote{https://lightedge.com/the-data-explosion-and-hidden-data-storage-costs-in-the-cloud-could-object-storage-be-the-answer/}%
\smallskip
\end{quote}

This tremendous increase in data usage has a substantial impact on businesses'
storage costs. A recent report by the online magazine Tech Monitor quotes a
survey of UK tech managers that \textquotedblleft more than half of UK IT
decision-makers say data storage costs are unsustainable.\textquotedblright%
\footnote{https://www.techmonitor.ai/technology/data/data-storage-costs-uk-it}
This is particularly true for firms that collect and store data for the
purpose of selling data access to users. The costs associated with keeping
data include power and cooling (storage hardware must be kept powered, cooled,
and available 24/7, even if the demand for data access is not continuous),
hardware refresh and depreciation (storage needs replacement just to hold data
safely), replication, redundancy and reliability (to store data safely, one
needs multiple copies and backups), and maintenance and operations (the
overhead of making sure stored data remains accessible, safe and compliant).
In 2024, the average annual cost to operate a large data center ranged from
\$10 million to \$25 million.\footnote{See
https://www.concretelogicpodcast.com/blog/the-economics-of-data-centers-a-deep-dive-into-costs-and-revenues/
and
https://www.hivenet.com/post/a-comprehensive-guide-to-datacenter-cost-management.}%

If data storage and maintenance is a scarce and costly resource, then its
allocation becomes an economic problem. How much data should society keep, and
which kinds of data should it dump? How should this decision reflect the
preferences of data users? Are there incentive issues that might distort the
decision? How would a profit-maximizing owner of proprietary data price and
allocate access to the stored data? This paper offers a simple theoretical
model that addresses these questions.

A model of data-storage management should articulate its scope by defining
three aspects: what the data is used for, what the data consists of, who
curates the data and controls its access, and what their motivation is.
Regarding aspect (1), demand for data in our model originates from users'
interest in training statistical models. Our data users do not seek
information about individuals; rather, they wish to learn parameters of some
population-level statistical model. Demand is differentiated because different
user types are interested in different parameters. In particular, users differ
in the \textit{specificity} of what they are trying to learn. For example, an
AI language model may be trained to \textquotedblleft
understand\textquotedblright\ general text corpora, or texts in a specific
professional domain such as math or medicine. Such models correspond to users
who try to learn statistical parameters of broad and narrow domains, respectively.

For concreteness, we adopt the following specification. There are two user
types: \textquotedblleft nowcasters\textquotedblright\ and \textquotedblleft
forecasters\textquotedblright. The former want to learn the \textit{current
value} of a \textit{parameter}, whereas the latter want to learn its
underlying \textit{long-run value}. We primarily think of this dichotomy as a
metaphor for the general distinction between users who try to learn parameters
of narrow vs. broad domains. However, it also has concrete economic
interpretations. For example, a business may be interested in consumer data
for the purposes of pricing an existing product or for designing a new
product; the former requires short-term diagnosis, while the latter requires
long-term predictions. Likewise, academic researchers demand data for
policy-oriented or basic research; the former is concerned with precise
short-term predictions, while the latter aims at learning long-term fundamentals.

Regarding aspect (2), a database in our model consists of two random samples
from two time periods: the present and the past. Each data point has both
time-specific and idiosyncratic noise components. Thus, all observations from
some time period share the same time-specific noise realization, while having
independent idiosyncratic noise realizations. The parameter and noise terms
are independent Gaussians. Each user type aims to minimize the mean squared
error of the prediction he is interested in. This objective function induces a
value that each user type attaches to a sample defined by the number of
historical and current observations.

As to aspect (3), data in our model is curated by a monopolistic,
profit-maximizing firm, which controls users' access to the data. This has
real-life analogues. Companies such as NielsenIQ provide access to exclusive
consumer data, partly for the purpose of general consumer research. A more
recent example is the collaboration between Getty Images and Defined.AI, which
involves creating exclusive datasets and providing access to them for training
visual AI models.\footnote{See https://en.wikipedia.org/wiki/NielsenIQ and
https://finance.yahoo.com/news/defined-ai-announces-strategic-engagement-183000929.html.}
Such exclusive ownership of data for AI training seems to be an emerging trend
in the information landscape.\footnote{See, e.g.,
https://www.forbes.com/sites/kolawolesamueladebayo/2025/02/25/why-proprietary-data-is-the-new-gold-for-ai-companies/}%

In our model, storing a data point has a constant marginal cost. This captures
the physical costs of data storage and maintenance. The firm chooses the total
size of its database and its composition, i.e. the fraction of current and
historical data. If the firm were perfectly informed of users' type, it would
offer all users full access to the data and charge each user his ex-ante
valuation of the information inherent in a sample of the given size and
composition. Since users' type is their private information, the firm offers a
menu of data-access plans. Each plan consists of a fee as well as a level of
access to the database's two components.

Our main results characterize the monopolist's optimal menu. We first
establish that nowcasters are like \textquotedblleft high\textquotedblright%
\ types in a standard second-degree price discrimination model: Their
willingness to pay for any sample always exceeds the forecasters'. However,
our user typology does $not$ satisfy a single-crossing property: The
difference between the two types' willingness to pay increases with the size
of the current sub-sample, but $decreases$ with the size of the historical sub-sample.

Using this characterization, we show that the optimal menu gives universal
access to historical data. Nowcasters get full access to current data as well.
When forecasters' fraction in the population is above some threshold, the menu
offers both types the same full-access plan. However, if forecasters' fraction
is below the threshold, they get $no$ access to current data, in return for a
lower fee. These features are reminiscent of some real-life payment plans for
financial data.\footnote{For financial data from the Gulf Mercantile Exchange,
the CME group charges a premium fee for real-time data and a lower fee for
historical data. A subscription for real-time data includes access to delayed
and historical information. See
https://www.cmegroup.com/market-data/files/january-2025-market-data-fee-list.pdf.
Similarly, at Interactive Brokers, an account comes with delayed market data
free of charge, but a paid subscription is required for real-time data. See
https://www.interactivebrokers.com/campus/trading-lessons/market-data-for-advisors/.}%

We complete this analysis by studying the distortions of the database size and
composition that arise from second-degree price discrimination. The historical
sub-sample is too large and the current sub-sample is too small, relative to
the social optimum. Indeed, we may end up having more historical than current
data, unlike the social optimum. As to the total size of the database, there
is no clear-cut comparison. We show numerically that the database may be
\textit{too large} relative to the social optimum. Thus, relying on users'
incentives to manage data access can give rise to insufficient data dumping.

Finally, we further explore the theme of \textquotedblleft
narrow\textquotedblright\ and \textquotedblleft broad\textquotedblright\ data
users as high- and low-valuation types, using an alternative state space. In
this variant, each type wishes to learn a specific convex combination of two
independently distributed parameters. A database consists of two independent
samples, one for each parameter. User types whose convexification parameter is
closer to $\frac{1}{2}$ are \textquotedblleft broad\textquotedblright\ in the
sense that they display similar interest in the two parameters. In contrast,
types whose convexification parameter is close to $1$ or $0$ are
\textquotedblleft narrow\textquotedblright\ in the sense of being more
single-minded. Narrow types have a higher valuation of symmetric databases
than broad types. We characterize the optimal menu when there are four user
types, symmetrically distributed around $\frac{1}{2}$. The narrow types get
full data access, whereas the broad types either get full access or no access
at all. The second-best database is symmetric and smaller than the social
optimum. This variant demonstrates that our narrow/broad typology of data
users may have broad relevance (no pun intended).

\section{The Model}

A monopolistic \textit{firm} designs a dataset and controls its access to
\textit{users}. The population of users has measure one. There are two types
of users: \textquotedblleft\textit{nowcasters}\textquotedblright\ (denoted
$S$) interested in short-term prediction, and \textquotedblleft%
\textit{forecasters}\textquotedblright\ (denoted $L$) interested in long-term
prediction. Let $\lambda\in\lbrack0,1]$ denote the fraction of type-$S$ users
in the population.

Let $\mu\sim N(0,\sigma_{\mu}^{2})$ be a fixed \textit{parameter} of interest.
There are two \textit{time periods}, denoted $1$ (\textquotedblleft the
present\textquotedblright) and $0$ (\textquotedblleft the
past\textquotedblright). A \textit{database} is described by a pair
$(n_{0},n_{1})$, where $n_{t}$ indicates the size of a \textit{sample}
consisting of observations from period $t$. Each observation $i=1,...,n_{t}$
from the period-$t$ sample is a realization%
\[
y_{t,i}=\mu+x_{t}+\varepsilon_{t,i}%
\]
where $x_{t}\sim N(0,1)$ and $\varepsilon_{t,i}\sim N(0,\sigma_{\varepsilon
}^{2})$. The variance of $x_{t}$ is a normalization that entails no loss of
generality. The value of $x_{t}$ is drawn independently for each period $t$,
but its value is the same for all observations that belong to the period-$t$
sample. The value of $\varepsilon_{t,i}$ is drawn independently for every
$t,i$. Each data point in the database carries a \textit{storage cost} of
$c>0$.

In the analysis, we will normalize $\sigma_{\varepsilon}^{2}=1$. Although we
have already normalized $Var(x_{t})$, this additional normalization is without
loss of generality, in the sense that we can regard it as a
\textit{redefinition} of the unit of measurement of database size: $n_{t}$ is
effectively measured in terms of multiples of $\sigma_{\varepsilon}^{2}$.

The two user types differ in what they try to learn. After learning from
whatever sample he gets access to, each type chooses an action $a\in%
\mathbb{R}
$. The two types' payoff functions are:%
\begin{align*}
u_{S}(a,\mu,x_{1})  &  =-(a-\mu-x_{1})^{2}\\
u_{L}(a,\mu)  &  =-(a-\mu)^{2}%
\end{align*}
The interpretation is that $\mu+x_{1}$ is the true \textit{current} value of a
variable of interest. Nowcasters, with their short-term prediction horizon,
try to learn this value. In comparison, $\mu$ is the variable's true
\textit{long-run} value. Forecasters, with their long-term prediction horizon,
try to learn this value.

The nowcaster/forecaster dichotomy captures the general distinction between
user types who differ in the \textit{specificity} of their learning objective.
Nowcasters narrowly focus on learning the parameter $\mu+x_{1}$ that governs
current data. Forecasters have a broader objective of learning the underlying
parameter $\mu$ that governs both present and historical data.

Users are Bayesian expected-utility maximizers. Type $k$'s willingness to pay
for access to a sample $(n_{0},n_{1})$, denoted $V_{k}(n_{0},n_{1})$, is equal
to the expected-utility gain that the information in the database generates.
We derive exact expressions for $V_{k}$ in Section 3. For analytical
convenience, we shall henceforth allow $n_{t}$ to take any non-negative real value.

A perfect monopolist can identify user types and extract their willingness to
pay. It is clear that users will receive full access, because their
willingness to pay is increasing in the amount of information provided.
Therefore, the perfect monopolist will choose the database $(n_{0},n_{1})$ to
solve the following maximization problem:%
\begin{equation}
\max_{n_{0},n_{1}}\text{ }\{\lambda V_{S}(n_{0},n_{1})+(1-\lambda)V_{L}%
(n_{0},n_{1})-c(n_{0}+n_{1})\} \label{FB problem}%
\end{equation}
We refer to a solution to this problem as the \textit{first-best solution}.

The main problem we analyze is based on the assumption that users' types are
their \textit{private information}. Consequently, applying the revelation
principle, the monopolist offers a menu $M$ of \textit{access plans}
$m^{k}=(q_{0}^{k},q_{1}^{k},p^{k})$, where $q_{t}^{k}\in\lbrack0,n_{t}]$
represents the level of access that user type $k$ gets to the period-$t$
sample, and $p^{k}\geq0$ is the fixed access fee he pays. The usual
participation and incentive-compatibility constraints must hold.

Thus, our monopolist's second-best maximization problem is%
\begin{equation}
\max_{n_{0},n_{1},(q_{0}^{k},q_{1}^{k},p^{k})_{k=S,L}}\text{ }\{\lambda
p^{S}+(1-\lambda)p^{L}-c(n_{0}+n_{1})\} \label{SB problem}%
\end{equation}
subject to the constraints%
\begin{align*}
n_{t}  &  \geq q_{t}^{k}\geq0\\
V_{k}(q_{0}^{k},q_{1}^{k})-p^{k}  &  \geq0\\
V_{k}(q_{0}^{k},q_{1}^{k})-p^{k}  &  \geq V_{k}(q_{0}^{-k},q_{1}^{-k})-p^{-k}%
\end{align*}
for every $t=0,1$ and every $k=S,L$ ($-k$ denotes the other user type).

The first constraint means that users get potentially partial access to the
database that the monopolist chooses to curate. The second constraint is user
type $k$'s participation (IR) constraint, and the third constraint is type
$k$'s incentive-compatibility (IC) constraint. We refer to a solution to
(\ref{SB problem}) as a \textit{second-best solution}.

The monopolist in our model chooses the size and composition of a database, as
well as how to price user access to the database. We regard the first
component as a \textquotedblleft data dumping\textquotedblright\ decision. Our
interpretation is that the monopolist controls an extremely large set of data
points from both time periods. The data is costly to store, and so the
monopolist has to decide how much data from each time period to keep, while
deleting the rest. It should be emphasized that storage costs are fixed and
depend on the database $(n_{0},n_{1})$, rather than on the access granted to
users as described by $q$.

\section{Preliminary Analysis: Value of Data}

In this section we derive formulas for users' willingness to pay for data
access, and highlight their key properties.

Let $\theta^{k}$ denote user type $k$'s \textquotedblleft
target\textquotedblright\ --- i.e., $\theta^{S}=\mu+x_{1}$ and $\theta^{L}%
=\mu$. Each type's prior belief over his target is Gaussian. Since signals are
Gaussian as well, so is each type's posterior belief. By standard arguments,
since a user's loss function is the squared error of his target prediction,
the user's willingness to pay for $(n_{0},n_{1})$ is equal to the
\textit{reduction in the variance} of his belief over his target. The prior
variances over $\theta^{S}$ and $\theta^{L}$ are $\sigma_{\mu}^{2}+1$ and
$\sigma_{\mu}^{2}$, respectively. Let us now calculate the variance of types'
posterior beliefs.

From type $L$'s point of view, a period-$t$ sample generates a signal $\bar
{y}_{t}=\theta^{L}+x_{t}+\bar{\varepsilon}_{t}$, where%
\[
\bar{\varepsilon}_{t}=\sum\frac{\varepsilon_{t,i}}{n_{t}}%
\]
is the average observational noise in the period-$t$ sample. The variance of
the period-$t$ signal conditional on $\theta^{L}$ is $1+\sigma_{\varepsilon
}^{2}/n_{t}$. From type $S$'s point of view, the two periods' samples generate
the signals $\bar{y}_{1}=\theta^{S}+\bar{\varepsilon}_{1}$ and $\bar{y}%
_{0}=\theta^{S}+x_{0}-x_{1}+\bar{\varepsilon}_{0}$, where $\bar{\varepsilon
}_{0}$ is defined as before. Note that unlike the case of type $L$, the error
term in $\bar{y}_{0}$ is not independent of $\theta^{S}$ because both include
$x_{1}$.

Applying the standard Gaussian signal extraction formula to the signals
provided by the two periods' samples, we obtain the following result.\bigskip

\begin{remark}
\label{prop VL VS}The user types' willingness to pay for $(n_{0},n_{1})$ is%
\begin{equation}
V_{L}(n_{0},n_{1})=\frac{\sigma_{\mu}^{4}(n_{1}+n_{0}+2n_{0}n_{1})}%
{\sigma_{\mu}^{2}(n_{1}+n_{0}+2n_{0}n_{1})+(1+n_{0})(1+n_{1})} \label{VL}%
\end{equation}%
\begin{equation}
V_{S}(n_{0},n_{1})=V_{L}(n_{0},n_{1})+\frac{3\sigma_{\mu}^{2}n_{0}%
n_{1}+2\sigma_{\mu}^{2}n_{1}+n_{1}+n_{0}n_{1}}{\sigma_{\mu}^{2}(n_{1}%
+n_{0}+2n_{0}n_{1})+(1+n_{0})(1+n_{1})}\bigskip\label{VS}%
\end{equation}

\end{remark}

These formulas have simple, interpretable and useful properties, which the
following result collects.\bigskip

\begin{remark}
\label{remark properties}The functions $V_{L}$ and $V_{S}$ satisfy the
following properties:\newline(i) $V_{L}$ and $V_{S}$ are strictly increasing
in both arguments.\newline(ii) $V_{L}$ and $V_{S}$ are strictly concave. In
particular, $\partial^{2}V_{k}(n_{0},n_{1})/\partial n_{t}^{2}$ and
$\partial^{2}V_{k}(n_{0},n_{1})/\partial n_{0}\partial n_{1}$ are strictly
negative for every type $k$ and period $t$.\newline(iii) $V_{L}$ is symmetric.
In contrast, for every $(n_{0},n_{1})$, $V_{S}(x,y)>V_{S}(y,x)$ if $y>x$, and%
\[
\frac{\partial V_{S}(n_{0},n_{1})}{\partial n_{1}}>\frac{\partial V_{S}%
(n_{0},n_{1})}{\partial n_{0}}%
\]
\newline(iv) $V_{S}(0,0)=V_{L}(0,0)=0$; $V_{S}(n_{0},0)=V_{L}(n_{0},0)$; and
$V_{S}(n_{0},n_{1})>V_{L}(n_{0},n_{1})$ for every $n_{0}\geq0$ and $n_{1}%
>0$.\newline(v) For every $(n_{0},n_{1})$,%
\begin{align*}
\frac{\partial V_{S}(n_{0},n_{1})}{\partial n_{1}}  &  >\frac{\partial
V_{L}(n_{0},n_{1})}{\partial n_{1}}\\
\frac{\partial V_{L}(n_{0},n_{1})}{\partial n_{0}}  &  >\frac{\partial
V_{S}(n_{0},n_{1})}{\partial n_{0}}%
\end{align*}
\bigskip
\end{remark}

The remarks' proofs are relegated to Appendix II. However, the intuition
behind the properties in Remark \ref{remark properties} is important for the
subsequent analysis. Parts $(i)$ and $(ii)$ are simple consequences of $V_{L}$
and $V_{S}$ being value-of-information functions. First, they are strictly
increasing in sample size because information always has positive marginal
value in this environment. Second, the functions are strictly concave because
information has diminishing marginal value in this environment: The marginal
variance reduction that an additional sample point from any period generates
gets smaller as we increase any period's sample size.

Part $(iii)$ articulates a difference in how the two types regard sample
points from each period. For type $L$, the two periods are symmetric: If we
permute $n_{0}$ and $n_{1}$, the sample is equally informative for this type.
In contrast, for type $S$, a present sample point is always more informative
than a historical sample point, because the latter has another layer of
independent noise (given by $x_{0}-x_{1}$) relative to the former. This is
unsurprising: A nowcaster,\ who is trying to learn something about the
present, will intuitively prefer a current observation to a historical one.

Part $(iv)$ means that type $S$ is a \textquotedblleft high\textquotedblright%
\ type relative to type $L$: His willingness to pay for non-null samples is
always strictly higher. For an intuition behind this result, imagine that
there is no observation-specific noise, such that the only sources of noise
are the time-specific\ random variables $x_{0}$ and $x_{1}$. For type $L$, the
two samples provide two independent signals of variance $1$ regarding his
target. In contrast, for type $S$, the current sample offers a
\textit{perfect} signal of his target, while the historical sample offers a
signal of variance $2$. Clearly, the database is more informative for type $S$
than for type $L$. This ranking continues to hold when we reintroduce the
independent, observation-specific noise. It reflects a general convexity
property: In the Gaussian context, having two independent signals of variance
$A$ and $B$ is more informative than having two independent signals of
variance $\frac{1}{2}(A+B)$ each.

As part $(v)$ articulates, the classification of the two types as
\textquotedblleft high\textquotedblright\ and \textquotedblleft
low\textquotedblright\ does not translate to a standard single-crossing
property w.r.t the natural partial ordering of pairs $(n_{0},n_{1})$. On one
hand, both $V_{L}$ and $V_{S}$ increase in this order (by part $(i)$ of the
remark). However, while an increase in $n_{1}$ leads to an increase in the
difference $V_{S}(n_{0},n_{1})-V_{L}(n_{0},n_{1})$ --- as a standard
single-crossing property would prescribe --- an increase in $n_{0}$ leads to a
\textit{decrease} in $V_{S}(n_{0},n_{1})-V_{L}(n_{0},n_{1})$, which goes
against the single-crossing property.

The fact that nowcasters value statistical data more than forecasters, coupled
with the two types' radically different marginal attitude to the two kinds of
statistical data, will drive our results in the next section.

\section{Main Results}

This section characterizes the monopolist's optimal policy, including the size
and composition of the database, the level of access offered to each user
type, and the structure of access fees.

As a benchmark, let us present the solution to the first-best problem
(\ref{FB problem}). Since $V_{L}$ and $V_{S}$ are strictly concave, the
optimal database $(n_{0}^{\ast},n_{1}^{\ast})$ is uniquely given by
first-order conditions:%
\begin{align}
(1-\lambda)\frac{\partial V_{L}(n_{0},n_{1})}{\partial n_{1}}+\lambda
\frac{\partial V_{S}(n_{0},n_{1})}{\partial n_{1}}  &  =c\label{FOC FB}\\
(1-\lambda)\frac{\partial V_{L}(n_{0},n_{1})}{\partial n_{0}}+\lambda
\frac{\partial V_{S}(n_{0},n_{1})}{\partial n_{0}}  &  =c\nonumber
\end{align}
whenever $n_{0}^{\ast},n_{1}^{\ast}>0$. Moreover, it is optimal for the firm
to offer users full access to the database, and charge each type $k$ his
willingness to pay $V_{k}(n_{0}^{\ast},n_{1}^{\ast})$ as an access fee. The
following result characterizes the composition of the optimal
database.\bigskip

\begin{remark}
\label{FB database}The optimal database $(n_{0}^{\ast},n_{1}^{\ast})$
satisfies $n_{1}^{\ast}\geq n_{0}^{\ast}$. Moreover, the inequality is strict
when $n_{0}^{\ast}>0$.\bigskip
\end{remark}

Thus, the first-best database contains more current data points than
historical ones.

We now turn to the second-best problem. Our first result characterizes the
optimal menu for a given database $(n_{0},n_{1})$. Denote%
\begin{equation}
\lambda^{\ast}(n_{0})=\left(  \frac{\sigma_{\mu}^{2}}{1+\sigma_{\mu}^{2}%
\frac{2n_{0}+1}{n_{0}+1}}\right)  ^{2}\label{lambda star}%
\end{equation}
Observe that%
\begin{equation}
\left(  \frac{\sigma_{\mu}^{2}}{1+2\sigma_{\mu}^{2}}\right)  ^{2}\leq
\lambda^{\ast}(n_{0})\leq\left(  \frac{\sigma_{\mu}^{2}}{1+\sigma_{\mu}^{2}%
}\right)  ^{2}\label{lambda bounds}%
\end{equation}
for every $n_{0}$. That is, $\lambda^{\ast}(n_{0})$ is interior and bounded
away from $0$ and $1$.\bigskip

\begin{proposition}
\label{Prop SB}For fixed $(n_{0},n_{1})$, the second-best solution has the
following properties:\newline(i) $q_{0}^{S}=q_{0}^{L}=n_{0}$.\newline(ii)
$q_{1}^{S}=n_{1}$; and $q_{1}^{L}=n_{1}$ if $\lambda\leq\lambda^{\ast}(n_{0}%
)$, while $q_{1}^{L}=0$ if $\lambda>\lambda^{\ast}(n_{0})$.\newline(iii) The
access fees paid by each user type are%
\begin{align*}
p^{L} &  =V_{L}(n_{0},q_{1}^{L})\\
p^{S} &  =V_{S}(n_{0},n_{1})-V_{S}(n_{0},q_{1}^{L})+V_{L}(n_{0},q_{1}^{L})
\end{align*}

\end{proposition}

Thus, optimal second-degree discrimination exhibits a \textquotedblleft
bang-bang\textquotedblright\ property. When the fraction of nowcasters in the
user population is below some threshold, there is no discrimination: Both
types are offered full access to the data for a uniform fee that extracts the
forecasters' entire surplus. When the fraction of nowcasters exceeds the
threshold, forecasters pay a relatively low fee and in return forego their
access to current data, while nowcasters pay a premium to get full access to
all data.\footnote{Our characterizations of the first-best and second-best
extend to a cost function that is increasing and convex in $n_{0}+n_{1}$.}

The structure of the second-best menu stems from the fact that $V_{L}$ and
$V_{S}$ are value-of-information functions. By Remark \ref{remark properties},
nowcasters and forecasters are effectively \textquotedblleft
high\textquotedblright\ and \textquotedblleft low\textquotedblright\ types.
Therefore, the former's IC constraint and the latter's IR constraint are the
relevant ones and both bind. The systematic violation of the single-crossing
property explains the structure of discriminatory data access. The difference
between the high and low types' valuations of data access $(q_{0},q_{1})$
increases with $q_{1}$ but decreases with $q_{0}$. As a result, there is no
discrimination in historical data access, whereas discriminatory access to
current data can arise. The latter's bang-bang structure (as described in part
$(ii)$ of Proposition \ref{Prop SB}) arises from the fact that the ratio
between the two types' marginal valuation of current-data access is
independent of its size.

We now turn to an analysis of an optimal second-best database $(n_{0}^{\prime
},n_{1}^{\prime})$. We begin by addressing the relation between between
$n_{0}^{\prime}$ and $n_{1}^{\prime}$ and its dependence on the type
distribution.\bigskip

\begin{proposition}
\label{prop statics} Suppose that $c$ is not too large, such that in optimum
$(n_{0},n_{1})\neq(0,0)$. (i) If $\lambda$ is sufficiently close to $1$, then
$n_{1}^{\prime}>n_{0}^{\prime}=0$ and $q^{L}\neq q^{S}$. (ii) If
$\lambda<\sigma_{\mu}^{4}/(1+2\sigma_{\mu}^{2})^{2}$, then $n_{0}^{\prime
}=n_{1}^{\prime}$ and $q^{L}=q^{S}$. (iii) When $(n_{0}^{\prime},n_{1}%
^{\prime})$ is interior and $q^{L}\neq q^{S}$, $n_{0}^{\prime}$ decreases in
$\lambda$ and $n_{1}^{\prime}$ increases in $\lambda$.\bigskip
\end{proposition}

Thus, when the type distribution consists mostly of nowcasters, the current
sample is larger than the historical sample in optimum, because nowcasters
find the former more informative. In contrast, when the fraction of nowcasters
is sufficiently low, there is no discrimination and the database is symmetric.
The reason is that in this case, access is priced according to the
willingness-to-pay of the forecasters, for whom historical and current data
are equally informative. In between, if the second-best solution is interior
and with discrimination, increasing the fraction of nowcasters results in a
smaller historical sample and a larger current sample.

While this collection of observations might give an impression that
$n_{1}^{\prime}$ is always weakly above $n_{0}^{\prime}$ (as in the
first-best), this turns out not to be the case. For some parameter values ---
e.g., when $c=0.3,$ $\sigma_{\mu}=1.2$ and $\lambda\in(0.25,0.3)$ --- we have
$n_{0}^{\prime}>n_{1}^{\prime}$, i.e., there are more historical data points
than current ones. The underlying reason is that the following result
demonstrates, the second-best database distorts the data composition in favor
of historical data. For brevity, we focus on the case in which the latter is
interior.\bigskip

\begin{proposition}
\label{Prop FB vs SB}Let $(n_{0}^{\ast},n_{1}^{\ast})$ and $\left(
n_{0}^{\prime},n_{1}^{\prime}\right)  $ be the first-best and second-best
databases, respectively. Suppose $n_{t}^{\ast}>0$ for both $t=0,1$. Then,
$n_{0}^{\prime}>n_{0}^{\ast}$ and $n_{1}^{\prime}<n_{1}^{\ast}.$\bigskip
\end{proposition}

Thus, relative to the social optimum, a profit-maximizing monopolist dumps too
much current data and too little historical data. For an intuition behind this
result, consider the first-best database $(n_{0}^{\ast},n_{1}^{\ast})$. Under
the first-best, both types get full data access. Now suppose that forecasters
are denied access to current data --- as they are under the second-best when
the fraction of nowcasters is sufficiently high. As a result, the marginal
aggregate value of current data goes down, while that of historical data goes
up. As a result, the second-best values of $n_{1}$ and $n_{0}$ will be
adjusted downward and upward, respectively.

As to the total second-best database size $n_{0}^{\prime}+n_{1}^{\prime}$, one
might expect that the cost of screening user types will lead to an efficiency
loss in the form of under-storage of data. It turns out that this is
\textit{not} the case: The comparison between the first-best and second-best
total database size is not clear-cut. For instance, $n_{0}^{\prime}%
+n_{1}^{\prime}>n_{0}^{\ast}+n_{1}^{\ast}$ when $c=0.1$, $\sigma_{\mu}^{2}=2$,
and $\lambda>0.4$. A supplementary appendix provides numerical simulations
that demonstrate this effect for a range of parameter values.

The reason over-storage of data may arise in the second-best problem is that
to compensate for type $L$'s lack of access to current data, the firm inflates
the historical database. This increase may be so big that it outweighs the
reduction in the size of the current database. Thus, although incentive
constraints dissipate the value of available data for some users
(specifically, the forecasters who are interested in long-run predictions),
the monopolist's reaction to this effect can result in too little data dumping
relative to the social optimum.\bigskip

\noindent\textit{Comment on the model's temporal interpretation}

\noindent We have interpreted $n_{0}$ and $n_{1}$ as \textquotedblleft
old\textquotedblright\ and \textquotedblleft current\textquotedblright\ data.
If we think of the interaction between the monopolist and users as a one-off
event, this interpretation is airtight. Alternatively, suppose that the
monopolist is a long-run player interacting with a sequence of short-lived
users. At every period $t$, there is an arbitrarily large inflow of new
datapoints, and the monopolist decides how many of them to curate in the
\textquotedblleft current\textquotedblright\ database as well as how many of
the previously stored datapoints to dump. Datapoints that are more than
two-periods old are eliminated automatically. The pair $(n_{0},n_{1})$
represents a stationary data policy: At every $t$, $n_{0}$ $\;\not (
\;$$n_{1}$) represents the amount of data from period $t-1$ ($t$) that is kept
in the database.

Under this \textquotedblleft Markovian\textquotedblright\ interpretation,
\textquotedblleft current\textquotedblright\ data at period $t$ become
\textquotedblleft old\textquotedblright\ data at period $t+1$, hence
$n_{0}\leq n_{1}$. While this property holds anyway in our model under the
first-best solution, the second-best solution can violate it. Therefore, if we
want our model to be consistent with the Markovian interpretation, we should
add the constraint $n_{0}\leq n_{1}$ to the second-best problem.

As already noted, our model also has non-temporal interpretations, by which
$n_{1}$ and $n_{0}$ represent databases that belong to narrow and broad
domains, respectively. Such interpretations do not pose the problem discussed here.

\section{An Alternative Typology of Data Users}

In this section, we consider an environment with a different typology of
\textquotedblleft narrow\textquotedblright\ and \textquotedblleft
broad\textquotedblright\ users. We present it to demonstrate the broader
relevance of this distinction beyond the main model's specification, as well
as to further explore the incentive-compatibility constraints that arise when
users with different interests seek access to a common database.

For each $i\in\{0,1\}$, $\theta_{i}$ is an unknown constant that users wish to
learn. The prior distribution of $\theta_{i}$ is $N(0,1)$, independently
across $i$. A user type is defined by $\beta\in\lbrack0,1]$. As in the main
model, a user chooses an action $a$ after observing a sample. His objective is
to minimize the quadratic loss function%
\[
\lbrack a-(\beta\theta_{0}+(1-\beta)\theta_{1})]^{2}%
\]

One interpretation is that $\theta_{0}$ and $\theta_{1}$ are parameters that
characterize two different demographics (young vs. old, liberals vs.
conservatives, etc.). A user is interested in learning about a particular
group (a target audience for some product, an electoral district, etc.), which
is composed of different proportions of the two demographics. Thus, if
$\beta>\beta^{\prime}>\frac{1}{2}$ or $\beta<\beta^{\prime}<\frac{1}{2}$, we
can classify types $\beta$ and $\beta^{\prime}$ as \textquotedblleft
narrow\textquotedblright\ and \textquotedblleft broad\textquotedblright,
respectively, because type $\beta$ is relatively single-minded whereas type
$\beta^{\prime}$ has relatively balanced interests.

As in our main model, the monopolist commits to a pair $(n_{0},n_{1}%
)\geq(0,0)$ such that $n_{0}+n_{1}=n$, where $n_{i}$ is the size of a sample
of independent observations of $y_{i}=\theta_{i}+\varepsilon$, where
$\varepsilon\sim N(0,\sigma^{2})$ is independently distributed noise. Let
$\bar{y}_{i}$ denote the sample average for $i$ (given the sample size $n_{i}%
$). Then, $\bar{y}_{i}\sim N(\theta_{i},\frac{\sigma^{2}}{n_{i}})$. A user
evaluates a database $(n_{0},n_{1})$ by the expected-loss reduction it
generates.\bigskip

\begin{remark}
\label{remark beta}User $\beta$'s willingness to pay for the database
$(n_{0},n_{1})$ is%
\begin{equation}
U_{\beta}(n_{0},n_{1})\equiv\beta^{2}\frac{n_{0}}{n_{0}+\sigma^{2}}%
+(1-\beta)^{2}\frac{n_{1}}{n_{1}+\sigma^{2}} \label{U beta}%
\end{equation}

\end{remark}

A key property of (\ref{U beta}) is that when $n_{1}=n_{0}$, it attains a
minimum at $\beta=\frac{1}{2}$. In this sense, \textquotedblleft
narrow\textquotedblright\ and \textquotedblleft broad\textquotedblright\ types
are \textquotedblleft high valuation\textquotedblright\ and \textquotedblleft
low valuation\textquotedblright\ types, respectively.

From now on, we assume that the set of user types is $B=\{r,s,1-s,1-r\}$,
where $0<r<s<\frac{1}{2}$. The type distribution $p$ is symmetric:
$p(\beta)=p(1-\beta)$ for every $\beta\in B$. As in the main model, this type
space violates the single-crossing property. Although there are four types,
this specification is simpler than the one in the main model, for two reasons.
First, the value of data access is separable in the two data components
because they offer independent information about the two independently
distributed constants. Second, the symmetry of the type distribution
simplifies the characterization of binding constraints in the monopolist's problem.

Let us begin with the first-best. As usual, all types get full data access.
The database $(n_{0},n_{1})$ is chosen to maximize%
\[
\sum_{\beta}p(\beta)\left[  \beta^{2}\frac{n_{0}}{n_{0}+\sigma^{2}}%
+(1-\beta)^{2}\frac{n_{1}}{n_{1}+\sigma^{2}}\right]  -c(n_{0}+n_{1})
\]
The objective function is strictly concave, as well as separable and symmetric
in $n_{0}$ and $n_{1}$. Therefore, the optimal solution is symmetric,
$n_{0}^{\ast}=n_{1}^{\ast}=N^{\ast}$, and given by first-order conditions as
long as $c$ is not too large.

We now turn to the second-best problem. The monopolist commits to a database
$(n_{0},n_{1})\ $and a menu of access bundles and fees $\left\{  (q_{0}%
(\beta),q_{1}(\beta),t(\beta)\right\}  _{\beta\in B}$, to maximize%
\[
\sum_{\beta\in B}p(\beta)t(\beta)-c(n_{0}+n_{1})
\]
subject to the feasibility constraints, $0\leq q_{i}(\beta)\leq n_{i}$ for all
$\beta\in B$, as well as all participation and incentive-compatibility
constraints.\bigskip

\begin{proposition}
\label{prop beta second best}The monopolist's second-best solution is
characterized as follows:\newline(i) $n_{0}^{\ast\ast}=n_{1}^{\ast\ast
}=N^{\ast\ast}<N^{\ast}$.\newline(ii) $q_{i}^{\ast\ast}(r)=q_{i}^{\ast\ast
}(1-r)=N^{\ast\ast}$ for every $i=0,1$.\newline(i) For every type $\beta
\in\{s,1-s\}$ and every $i=0,1$,%
\[
q_{i}^{\ast\ast}(\beta)=\left\{
\begin{array}
[c]{ccc}%
N^{\ast\ast} & if & 2p(r)\leq\frac{s^{2}+(1-s)^{2}}{r^{2}+(1-r)^{2}}\\
0 & otherwise &
\end{array}
\right.
\]

\end{proposition}

This characterization is consistent with the main model's theme that narrow
types have high willingness to pay and therefore receive full data access,
while broad types have low willingness to pay and may therefore get restricted
data access. Here, broad types either get full or no access, whereas in the
main model the broad type was given full access to one dataset and potentially
no access to the other. This difference arises from the symmetries of the
present specification. It also leads to the clear-cut conclusion that the size
of both components of the second-best database is unambiguously \textit{below}
their first-best level.

\section{Related Literature}

Computer scientists have begun addressing the data-storage challenge in the
age of big data (e.g., Milo (2019), and Davidson et al. (2023)), by attempting
to develop effective and computationally efficient algorithms for determining
which pieces of data to delete. For examples of recent attempts to quantify
the cost of training AI models (which is partly a function of training-set
size), see Guerra et al. (2023) and Cottier et al. (2024).

Within economic theory and IO, our paper is closest to the growing literature
on markets for data and computing services. To our knowledge, our focus on the
data-dumping problem is new to this literature. Bergemann, Bonatti and Smolin
(2018) analyze a monopolist who designs a menu of Blackwell experiments for
buyers with the same state-dependent utility, but whose private type is their
prior belief over the states. Bergemann, Bonatti and Smolin (2025) consider a
monopolist who offers LLM models, which are used to compute a set of tasks. An
LLM model is characterized by three quantities that are costly to supply: the
number of input and output tokens to be allocated between the tasks, and the
number of \textquotedblleft fine tuning\textquotedblright\ tokens that can
improve the model's overall performance. The paper analyzes the design of the
optimal menu of LLM models for a population of consumers whose private types
are the weights they assign to each of the tasks. A similar typology of
consumers appears in Bergemann and Deb (2025), who consider monopolistic
screening of consumers who use a cloud service for a set of tasks.

A feature that is common to these papers and ours is the idea that buyers
require data to perform certain tasks. One crucial difference is the
public-good aspect of our model: Our monopolist first needs to determine the
size and composition of the data to hold, and screening is then done by
offering different degrees of access to this data for different fees. A second
major distinction is the buyer typology: Buyers differ in the parameters they
seek to infer from the dataset. This implies a difference in their
willingness-to-pay for data.

One strand of the literature on information markets focuses on buying and
selling of personal data, mainly for personalized advertising and price
discrimination. A notable paper that belongs to this strand is Bergemann and
Bonatti (2015), who study a market in which firms face uncertainty about the
match value with different consumer types. If a firm knew its match value with
a particular consumer type, it could optimally choose a marketing campaign
targeting that type. Firms can purchase this information from a single data
provider who charges a price per \textquotedblleft match-value
query\textquotedblright, which provides a firm with the set of consumer types
who generate the requested match value. The authors characterize firms' query
demand and the data provider's price-setting decision. Unlike our setting,
data buyers in Bergemann and Bonatti (2015) have no private information, hence
the monopolist does not face a screening problem. See Bergemann and Bonatti
(2019) for an extensive review of related works.

Another strand in this literature studies intermediaries who can provide hard
evidence on the quality of a product, whose provider can then decide whether
to disclose the evidence (see Ali et al. (2022) and the references therein).
By contrast, our focus is on the use of statistical data for general (i.e.,
not individual-specific) predictions. Such usage of data was recently studied
in a different context by Gans (2024): He asked whether users who freely
contribute training data for generative AI may have an incentive to stop doing
so once they start relying on the AI.

Our model is an example of monopolistic pricing of excludable public goods
(Brito and Oakland (1980), Norman (2004)). What is new here is that the public
good in our model is statistical data. It has two dimensions (historical and
current data), and users' demand for it originates from the informational
value of statistical data, which generates a structured violation of the
single-crossing property.

As an example of a two-type monopolistic screening problem without single
crossing, our paper is also related to Haghpanah and Siegel (2025), who
characterize the optimal menu for this case. A key lemma in that paper
concerns the notion of an \textquotedblleft uncontested type\textquotedblright%
, whose willingness-to-pay for his efficient option is higher than the other
type's willingness-to-pay for that same option. In an optimal menu, an
uncontested type is allocated his efficient outcome, while the other type's IR
constraint binds. In our framework, the monopolist first needs to decide on
the size and composition of the entire database. Given this database, all
costs are sunk and it is efficient to give everyone full data access.

Thus, one implication of Haghpanah and Siegel (2025) for our model is that
\textit{if} there is a high type, then given a database, that type will get
full access and the low type will have no surplus. Their model is silent on
the question of the optimal size and composition of the database. However,
once we fix a database and have identified that the nowcaster is the high
type, Haghpanah and Siegel (2025) provide a characterization of the
alternatives that are assigned to each type, in terms of new notions that they
define. Mapping these notions to our framework is more involved and less
transparent than giving a direct characterization of the optimal menu, as we
did here. Finally, since the Haghpanah-Siegel methods apply to two-type
environments, they are inapplicable to the model of Section 5.

\section{Conclusion}

This paper analyzed the management of anonymous statistical data as a
mechanism-design problem of allocating access to a non-rival public good,
taking into account fixed data management costs. Our focus here was on data
storage costs, but in general one could study variants on our model that
involve other data-management activities (such as data processing, which would
imply variable rather than fixed costs). By treating the allocation of access
to statistical data as a non-rival public good provision problem, our paper
naturally opens the door for follow-up questions: How would a benevolent
social planner price data access? How would the data-management industry
operate under different market structures? What would be the implications for
the regulation of this industry? Finally, the data storage dilemma is also
relevant when data users have other motivations, such as retrieving their
personal memories. In future work, we plan to develop an economic approach to
the data dumping problem for such environments.

\noindent{\LARGE Appendix I: Proofs}\bigskip

\noindent{\large Remark \textbf{\ref{FB database}}\medskip}

\noindent Suppose $n_{0}^{\ast}>n_{1}^{\ast}$. Suppose the firm deviates to
$(n_{0}^{\prime},n_{1}^{\prime})$ such that $n_{0}^{\prime}=n_{1}^{\ast}$ and
$n_{1}^{\prime}=n_{0}^{\ast}$. By Remark \ref{remark properties}(iii),
$V_{L}(n_{0}^{\prime},n_{1}^{\prime})=V_{L}(n_{0}^{\ast},n_{1}^{\ast})$,
whereas $V_{S}(n_{0}^{\prime},n_{1}^{\prime})>V_{S}(n_{0}^{\ast},n_{1}^{\ast
})$. Obviously, $c(n_{0}^{\prime}+n_{1}^{\prime})=c(n_{0}^{\ast}+n_{1}^{\ast
})$. Therefore, the deviation increases the value of the objective function
given by (\ref{FB problem}).

Now suppose $n_{0}^{\ast}=n_{1}^{\ast}>0$. Then, the optimum is given by
(\ref{FOC FB}). By Remark \ref{remark properties}(iii),%
\begin{align*}
\frac{\partial V_{L}(n_{0}^{\ast},n_{1}^{\ast})}{\partial n_{1}}  &
=\frac{\partial V_{L}(n_{0}^{\ast},n_{1}^{\ast})}{\partial n_{0}}\\
\frac{\partial V_{S}(n_{0}^{\ast},n_{1}^{\ast})}{\partial n_{1}}  &
>\frac{\partial V_{S}(n_{0}^{\ast},n_{1}^{\ast})}{\partial n_{0}}%
\end{align*}
contradicting (\ref{FOC FB}). $\blacksquare$\bigskip

\noindent{\large Proposition \textbf{\ref{Prop SB}}\medskip}

\noindent Fix $(n_{0},n_{1})$. We first solve the relaxed problem in which we
impose only three constraints: $IR_{L}$, $IC_{S}$ and the feasibility
constraint $n_{t}\geq q_{t}^{k}\geq0$ for $t=0,1$ and $k=L,S$. In the relaxed
problem, $IR_{L}$ and $IC_{S}$ must bind: If $IR_{L}$ has slack, the
monopolist can increase $p^{L}$ and $p^{S}$ by the same amount without
violating $IC_{S}$; and if $IC_{S}$ has slack, the monopolist can increase
$p^{S}$ without violating $IR_{L}.$

Define the function%
\[
\Delta\left(  q_{0},q_{1}\right)  :=V_{S}\left(  q_{0},q_{1}\right)
-V_{L}\left(  q_{0},q_{1}\right)
\]
This is the difference between the two types' willingness to pay. Substituting
the binding constraints into the monopolist's objective function reduces the
relaxed problem to the following optimization problem:%
\begin{equation}
\max_{(q_{0}^{L},q_{1}^{L},q_{0}^{S},q_{1}^{S})}\lambda\left[  V_{S}(q_{0}%
^{S},q_{1}^{S})-\Delta\left(  q_{0}^{L},q_{1}^{L}\right)  \right]  +\left(
1-\lambda\right)  V_{L}(q_{0}^{L},q_{1}^{L}) \label{relaxed problem proof}%
\end{equation}
subject to $n_{t}\geq q_{t}^{k}\geq0$ for $t=0,1$ and $k=L,S.$ Since $V_{S}$
increases in both its arguments, $(q_{0}^{S},q_{1}^{S})=(n_{0},n_{1})$ in
optimum. Since $\Delta\left(  q_{0}^{L},q_{1}^{L}\right)  $ decreases in
$q_{0}^{L}$ and $V_{L}$ increases in $q_{0}^{L}$, we must have $q_{0}%
^{L}=n_{0}$ in optimum. For $IR_{L}$ and $IC_{S}$ to bind, $p^{L}=V_{L}%
(n_{0},q_{1}^{L})$ and $p^{S}=V_{S}(n_{0},n_{1})-\Delta(n_{0},q_{1}^{L}).$ It
is now easy to verify that these prices, together with the allocation
identified above, satisfy the remaining constraints: $IR_{S}$ and $IC_{L}$.
Therefore, it is also a solution to (\ref{SB problem}) for the fixed
$(n_{0},n_{1})$.

From (\ref{VL}) and (\ref{VS}), it follows that%
\[
\frac{\partial V_{L}(q_{0}^{L},q_{1}^{L})/\partial q_{1}^{L}}{\partial
V_{S}(q_{0}^{L},q_{1}^{L})/\partial q_{1}^{L}}=\sigma_{\mu}^{4}\left(
\frac{q_{0}^{L}+1}{q_{0}^{L}+1+\sigma_{\mu}^{2}(2q_{0}^{L}+1)}\right)
^{2}=\lambda^{\ast}(q_{0}^{L})
\]
where $\lambda^{\ast}$ is as defined given by (\ref{lambda star}). We have
seen that $q_{0}^{L}=n_{0}$. Note that $\lambda^{\ast}(n_{0})$ is independent
of $q_{1}^{L}$. It follows that the derivative of the objective function in
the relaxed problem is positive for $\lambda<\lambda^{\ast}(n_{0})$ and
negative for $\lambda>\lambda^{\ast}(n_{0})$. This means that the optimal
solution for $q_{1}^{L}$ is extreme: $q_{1}^{L}=n_{1}$ for $\lambda
<\lambda^{\ast}(n_{0})$, and $q_{1}^{L}=0$ for $\lambda>\lambda^{\ast}(n_{0}%
)$. $\blacksquare\bigskip$

\noindent{\large Proposition \textbf{\ref{prop statics}}\medskip}

\noindent$(i)$ By Proposition \ref{Prop SB} and (\ref{lambda star}%
)-(\ref{lambda bounds}), for $\lambda>\sigma_{\mu}^{4}/(1+\sigma_{\mu}%
^{2})^{2}$, the second-best values of $q$ are $q^{L}=(n_{0},0)$ and
$q^{S}=(n_{0},n_{1})$. Plugging these values in (\ref{relaxed problem proof}),
and using the fact that $V_{S}(n_{0},0)=V_{L}(n_{0},0)$, the monopolist's
second-best payoff for given $(n_{0},n_{1})$ is%
\begin{equation}
\lambda V_{S}(n_{0},n_{1})+(1-\lambda)V_{L}(n_{0},0)-c(n_{0}+n_{1})
\label{second best discrimination payoff}%
\end{equation}
Property $(iii)$ in Remark \ref{remark properties} states that $\partial
V_{S}(n_{0},n_{1})/\partial n_{1}>\partial V_{S}(n_{0},n_{1})/\partial n_{0}$.
Moreover, the derivatives of $V_{S}$ and $V_{L}$ w.r.t $n_{0}$ and $n_{1}$ are
all continuous. Therefore, for $\lambda$ sufficiently close to one, $n_{0}=0$
in optimum, whereas $n_{1}>0$ is $c$ is not too large.\bigskip

\noindent$(ii)$ By Proposition \ref{Prop SB}, for $\lambda<\sigma_{\mu}%
^{4}/(1+2\sigma_{\mu}^{2})^{2}$, $q^{L}=q^{S}=(n_{0},n_{1})$. The monopolist's
second-best payoff thus becomes%
\[
V_{L}(n_{0},n_{1})-c(n_{0}+n_{1})
\]
Since $V_{L}$ is symmetric and strictly concave, the optimal solution
satisfies $n_{0}=n_{1}$.\bigskip

\noindent$(iii)$ When the second-best solution is interior and exhibits
discrimination between the two types, the optimal $(n_{0},n_{1})$ is derived
by applying first-order conditions to the strictly concave function
(\ref{second best discrimination payoff}):%
\begin{align*}
\lambda\frac{\partial V_{S}(n_{0},n_{1})}{\partial n_{1}}  &  =c\\
(1-\lambda)\frac{\partial V_{L}(n_{0},0)}{\partial n_{0}}+\lambda
\frac{\partial V_{S}(n_{0},n_{1})}{\partial n_{0}}  &  =c
\end{align*}
which by (\ref{VL}) and (\ref{VS}) are given by%
\begin{equation}
\frac{\lambda\left(  n_{0}+\sigma_{\mu}^{2}+2n_{0}\sigma_{\mu}^{2}+1\right)
^{2}}{\left(  n_{0}+n_{1}+n_{0}\sigma_{\mu}^{2}+n_{1}\sigma_{\mu}^{2}%
+n_{0}n_{1}+2n_{0}n_{1}\sigma_{\mu}^{2}+1\right)  ^{2}}=c \label{SB FOC n1}%
\end{equation}
and%
\begin{equation}
\frac{\lambda\sigma_{\mu}^{4}}{\left(  n_{0}+n_{1}+n_{0}\sigma_{\mu}^{2}%
+n_{1}\sigma_{\mu}^{2}+n_{0}n_{1}+2n_{0}n_{1}\sigma_{\mu}^{2}+1\right)  ^{2}%
}+\frac{(1-\lambda)\sigma_{\mu}^{4}}{\left(  n_{0}\sigma_{\mu}^{2}%
+n_{0}+1\right)  ^{2}}=c \label{SB (FOC n0)}%
\end{equation}
From these equations it follows that%
\begin{equation}
n_{1}=\sqrt{\frac{\lambda}{c}}-\frac{n_{0}+n_{0}\sigma_{\mu}^{2}+1}%
{n_{0}+\sigma_{\mu}^{2}+2n_{0}\sigma_{\mu}^{2}+1} \label{n1 SB}%
\end{equation}
Differentiating the R.H.S. w.r.t $n_{0}$, we can see that as $n_{0}$
\textit{decreases}, $n_{1}$ \textit{increases}. Thus, if $n_{0}$ decreases
when $\lambda$ increases, then whenever $n_{1}>n_{0}$ for some $\left(
\lambda,\sigma_{\mu},c\right)  ,$ this continues to be true for $\lambda
^{\prime}>\lambda.$

We now show that indeed, $n_{0}$ decreases in $\lambda.$ Plugging equation
(\ref{n1 SB}) into equation (\ref{SB (FOC n0)}) and rearranging yields
\[
(1-\lambda)\frac{\sigma_{\mu}^{4}}{\left(  n_{0}\sigma_{\mu}^{2}%
+n_{0}+1\right)  ^{2}}+c\frac{\sigma_{\mu}^{4}}{\left(  n_{0}+\sigma_{\mu}%
^{2}+2n_{0}\sigma_{\mu}^{2}+1\right)  ^{2}}=c
\]
Note that the L.H.S. of this equation decreases in $\lambda$ and also
decreases in $n_{0}.$ Hence, if $\lambda$ increases, $n_{0}$ decreases and so
$n_{1}$ increases. $\blacksquare$\bigskip

\noindent{\large Proposition }\textbf{\ref{Prop FB vs SB}}\medskip

\noindent Throughout this proof, we take it as given that the first-best and
second-best databases are strictly positive.

Let $f_{1}\left(  n_{0};x\right)  $ be a function that maps each value of
$n_{0}$ to a value of $n_{1}$ that solves the equation,%
\begin{equation}
(1-\lambda)\frac{\partial V_{L}(n_{0},n_{1})}{\partial n_{1}}+\lambda
\frac{\partial V_{S}(n_{0},n_{1})}{\partial n_{1}}=x \label{FB FOC n1}%
\end{equation}
Likewise, let $f_{0}\left(  n_{0};y\right)  $ be a function that maps each
value of $n_{0}$ to a value of $n_{1}$ that solves the equation,%
\begin{equation}
(1-\lambda)\frac{\partial V_{L}(n_{0},n_{1})}{\partial n_{0}}+\lambda
\frac{\partial V_{S}(n_{0},n_{1})}{\partial n_{0}}=y \label{FB FOC n0}%
\end{equation}
The L.H.S. of equations (\ref{FB FOC n1}) and (\ref{FB FOC n0}) are the
derivatives of the first-best objective function (\ref{FB problem}) w.r.t
$n_{1}$ and $n_{0}$, respectively.

We can regard $f_{0}\left(  n_{0};x\right)  $ and $f_{1}\left(  n_{0}%
;y\right)  $ as downward-sloping \textquotedblleft iso-marginal
value\textquotedblright\ curves in the space $%
\mathbb{R}
_{+}^{2}$, where the horizontal and vertical axes represent $n_{0}$ and
$n_{1}$, respectively, as in Figure 1.

We first establish that the curve that represents $f_{1}\left(  n_{0}%
;c\right)  $ intersects the curve that represents $f_{0}\left(  n_{0}%
;c\right)  $ from below at a single point $(n_{0}^{\ast},n_{1}^{\ast})$. By
part $(ii)$ of Remark \ref{remark properties}, $f_{0}\left(  n_{0};x\right)  $
and $f_{1}\left(  n_{0};y\right)  $ are both decreasing in $n_{0}$ for every
$x$ and $y$, and there is a unique pair $(n_{0}^{\ast},n_{1}^{\ast})$ (the
unique solution to the first-best problem, which is interior by assumption)
satisfying $n_{1}^{\ast}=f_{1}\left(  n_{0}^{\ast};c\right)  =f_{0}\left(
n_{0}^{\ast};c\right)  .$ We claim that $f_{1}\left(  n_{0};c\right)
<f_{0}\left(  n_{0};c\right)  $ for $n_{0}<n_{0}^{\ast}$ and $f_{1}\left(
n_{0};c\right)  >f_{0}\left(  n_{0};c\right)  $ for $n_{0}>n_{0}^{\ast}$. To
see why, recall that $n_{1}^{\ast}>n_{0}^{\ast}$ for all $\lambda>0$. By part
$(iii)$ of Remark \ref{remark properties}, for $n_{0}=n_{1}=a$ satisfying
$a=f_{1}\left(  a;c\right)  $ we have $a=f_{0}\left(  a;c^{\prime}\right)  $
for some $c^{\prime}<c.$ Hence, by part $(ii)$ of Remark
\ref{remark properties}, $f_{0}\left(  a;c\right)  <a.$ Since there is a
unique solution to $f_{1}\left(  n_{0}^{\ast};c\right)  =f_{0}\left(
n_{0}^{\ast};c\right)  ,$ it follows that $f_{1}\left(  n_{0};c\right)
>f_{0}\left(  n_{0};c\right)  $ for $n_{0}>n_{0}^{\ast}$ while $f_{1}\left(
n_{0};c\right)  <f_{0}\left(  n_{0};c\right)  $ for $n_{0}<n_{0}^{\ast}$.

We now argue that the second-best database $(n_{0}^{\prime},n_{1}^{\prime})$
satisfies $n_{0}^{\prime}>n_{0}^{\ast}$ and $n_{1}^{\prime}<n_{1}^{\ast}$ when
the second-best solution satisfies $q_{1}^{L}=0.$ To see this, let
$g_{1}(n_{0};x)$ and $g_{0}(n_{0};y)$ be the functions that map each value of
$n_{0}$ to the values of $n_{1}$ that solve the equations%
\begin{equation}
\lambda\frac{\partial V_{S}(n_{0},n_{1})}{\partial n_{1}}=x
\label{SB derivative n1}%
\end{equation}
and%
\begin{equation}
(1-\lambda)\frac{\partial V_{L}(n_{0},0)}{\partial n_{0}}+\lambda
\frac{\partial V_{S}(n_{0},n_{1})}{\partial n_{0}}=y \label{SB derivative n0}%
\end{equation}
respectively. The L.H.S. of equations (\ref{SB derivative n1}) and
(\ref{SB derivative n0}) are the derivatives of the relaxed second-best
objective function (\ref{second best discrimination payoff}) w.r.t $n_{1}$ and
$n_{0}$, respectively. By part $(ii)$ of Remark \ref{remark properties}, both
$g_{1}(n_{0};x)$ and $g_{0}(n_{0};y)$ are decreasing in $n_{0}.$ Thus, both
are represented by downward-sloping \textquotedblleft iso-marginal
value\textquotedblright\ curves in the same $%
\mathbb{R}
_{++}^{2}$ space we used to represent $f_{0}\left(  n_{0};x\right)  $ and
$f_{1}\left(  n_{0};y\right)  $. As we saw in the proof of Proposition
\ref{Prop SB}, there exists a unique $\left(  n_{0}^{\prime},n_{1}^{\prime
}\right)  $ satisfying $n_{1}^{\prime}=g_{1}(n_{0}^{\prime};c)=g_{0}%
(n_{0}^{\prime};c).$ We will now show that $n_{0}^{\prime}>n_{0}^{\ast}$ and
$n_{1}^{\prime}<n_{1}^{\ast}$ when $q_{1}^{L}=0.$

For any $(n_{0},n_{1})$, the L.H.S. of (\ref{SB derivative n1}) is lower than
the L.H.S. of (\ref{FB FOC n1}). By part $(ii)$ of Remark
\ref{remark properties}, $\frac{\partial}{\partial n_{1}}V_{S}(n_{0},n_{1})$
is decreasing in $n_{1}$. Therefore, the curve that represents $g_{1}%
(n_{0};x)$ lies \textit{below} the curve that represents $f_{1}(n_{0};x)$. In
a similar vein, part $(ii)$ of Remark \ref{remark properties} implies that
$\frac{\partial}{\partial n_{0}}V_{L}(n_{0},n_{1})<\frac{\partial}{\partial
n_{0}}V_{L}(n_{0},0)$, such that the L.H.S. of (\ref{SB derivative n0}) is
higher than the L.H.S. of (\ref{FB FOC n0}). Since $\frac{\partial}{\partial
n_{1}}V_{S}(n_{0},n_{1})$ is decreasing in $n_{1}$, it follows that the
iso-marginal value curve that represents $g_{0}(n_{0};x)$ lies \textit{above}
the curve that represents $f_{0}(n_{0};x)$. Figure 1 illustrates these curve
shifts.\bigskip%
\[
\text{Insert Figure 1}\bigskip
\]

As a result of the directions in which the curves that represent $g_{1}%
(n_{0};x)$ and $g_{0}(n_{0};y)$ are shifted relative to the curves that
represent $f_{1}(n_{0};x)$ and $f_{0}(n_{0};y)$, the unique intersection
$(n_{0}^{\prime},n_{1}^{\prime})$ of the curves that represent $g_{1}%
(n_{0};c)$ and $g_{0}(n_{0};c)$ satisfies $n_{0}^{\prime}>n_{0}^{\ast}$ and
$n_{1}^{\ast}>n_{1}^{\prime}$.

We next show that $n_{0}^{\prime}>n_{0}^{\ast}$ and $n_{1}^{\prime}%
<n_{1}^{\ast}$ also when $q_{1}^{L}=n_{1}.$ Recall that in this case, the
monopolist offers a single contract $(n_{0}^{\prime},n_{1}^{\prime},p)$, where
$p=V_{L}\left(  n_{0}^{\prime},n_{1}^{\prime}\right)  .$ Therefore, since
$V_{L}$ is strictly concave, $\left(  n_{0}^{\prime},n_{1}^{\prime}\right)  $
solve%
\begin{equation}
\frac{\partial V_{L}}{\partial n_{0}}(n_{0}^{\prime},n_{1}^{\prime}%
)=\frac{\partial V_{L}}{\partial n_{1}}(n_{0}^{\prime},n_{1}^{\prime})=c
\label{SB FOC q=n1}%
\end{equation}
By the symmetry of $V_{L}$, $n_{0}^{\prime}=n_{1}^{\prime}=b.$ We claim that
$n_{0}^{\ast}<b<n_{1}^{\ast}$. To see why, assume first that $b\geq
n_{1}^{\ast}$ (which implies that $b>n_{0}^{\ast}$ since $n_{1}^{\ast}%
>n_{0}^{\ast}$). Then,%
\[
c=(1-\lambda)\frac{\partial V_{L}(n_{0}^{\ast},n_{1}^{\ast})}{\partial n_{1}%
}+\lambda\frac{\partial V_{S}(n_{0}^{\ast},n_{1}^{\ast})}{\partial n_{1}%
}>\frac{\partial V_{L}(n_{0}^{\ast},n_{1}^{\ast})}{\partial n_{1}}%
>\frac{\partial V_{L}(b,b)}{\partial n_{1}}%
\]
where the first and second inequalities follow from parts $(v)$ and $(ii)$,
respectively, of Remark \ref{remark properties}. But the above inequality
violates equation (\ref{SB FOC q=n1}), a contradiction.

Next, assume $b\leq n_{0}^{\ast}$ (and hence, $b<n_{1}^{\ast}$). Then again by
Remark \ref{remark properties},
\[
c=(1-\lambda)\frac{\partial V_{L}(n_{0}^{\ast},n_{1}^{\ast})}{\partial n_{0}%
}+\lambda\frac{\partial V_{S}(n_{0}^{\ast},n_{1}^{\ast})}{\partial n_{0}%
}<\frac{\partial V_{L}(n_{0}^{\ast},n_{1}^{\ast})}{\partial n_{0}}%
<\frac{\partial V_{L}(b,b)}{\partial n_{0}}%
\]
violating equation (\ref{SB FOC q=n1}). $\blacksquare$\bigskip

\noindent{\large Remark }\textbf{\ref{remark beta}}\medskip

\noindent In the absence of any sample data, type $\beta$'s optimal action is
$a=E(\beta\theta_{0}+(1-\beta)\theta_{1})=0$. His prior expected loss is
$\beta^{2}+(1-\beta)^{2}$. To derive the user's expected loss given the
posterior belief induced by $(n_{0},n_{1})$, note that his action given the
sample outcome is%
\[
a=\beta E(\theta_{0}\mid\bar{y}_{0})+(1-\beta)E(\theta_{0}\mid\bar{y}_{1})
\]
This follows immediately from the quadratic loss function. Under the Gaussian
specification,%
\[
E(\theta_{x}\mid\bar{y}_{x})=\frac{1}{1+\frac{\sigma^{2}}{n_{x}}}\bar{y}_{x}%
\]
The user's ex-ante expected loss is%
\[
E\left[  \beta(E(\theta_{0}\mid\bar{y}_{0})-\theta_{0})+(1-\beta)(E(\theta
_{1}\mid\bar{y}_{1})-\theta_{1})\right]  ^{2}%
\]
where the ex-ante expectation is taken w.r.t. both the distributions over the
parameters $\theta_{0}$, $\theta_{1}$ and the sampling error $\varepsilon$.
Using the fact that these distributions are mutually independent, all have
mean zero, and implementing some algebra, we obtain that the expected loss is
given by%
\[
\beta^{2}\frac{\sigma^{2}}{n_{0}+\sigma^{2}}+(1-\beta)^{2}\frac{\sigma^{2}%
}{n_{1}+\sigma^{2}}%
\]
Hence, the expected loss reduction from $(n_{0},n_{1})$ is given by
(\ref{U beta}). $\blacksquare$\bigskip

\noindent{\large Proposition }\textbf{\ref{prop beta second best}}\medskip

\noindent For our purposes, we can consider a relaxed problem that only
examines a selection of the IR and IC\ constraints. The heuristic is to treat
$s$ and $1-s$ ($r$ and $1-r$ ) as \textquotedblleft low\textquotedblright%
\ (\textquotedblleft high\textquotedblright) types, such that the IR
constraints of $s$ and $1-s$, and the IC constraints of $r$ and $1-r$, should
be considered. Two additional IC\ constraints prevent types $s$ and $1-s$ from
mimicking each other. We will later verify that the other IR and IC
constraints can be ignored.

The following notation will be convenient:%
\[
x_{i}(\beta)=\frac{q_{i}(\beta)}{q_{i}(\beta)+\sigma^{2}}%
\]
Note that $x_{i}(\beta)$ is strictly increasing and strictly concave in
$q_{i}(\beta)$. We have the following six constraints in addition to the
feasibility constraints:%

\[%
\begin{array}
[c]{llll}%
{\small (1)} & {\small s}^{2}{\small x}_{0}{\small (s)+(1-s)}^{2}%
{\small x}_{1}{\small (s)-t(s)} & {\small \geq} & {\small 0}\\
{\small (2)} & {\small (1-s)}^{2}{\small x}_{0}{\small (1-s)+s}^{2}%
{\small x}_{1}{\small (1-s)-t(1-s)} & {\small \geq} & {\small 0}\\
{\small (3)} & {\small s}^{2}{\small x}_{0}{\small (s)+(1-s)}^{2}%
{\small x}_{1}{\small (s)-t(s)} & {\small \geq} & {\small s}^{2}{\small x}%
_{0}{\small (1-s)+(1-s)}^{2}{\small x}_{1}{\small (1-s)-t(1-s)}\\
{\small (4)} & {\small (1-s)}^{2}{\small x}_{0}{\small (1-s)+s}^{2}%
{\small x}_{1}{\small (1-s)-t(1-s)} & {\small \geq} & {\small (1-s)}%
^{2}{\small x}_{0}{\small (s)+s}^{2}{\small x}_{1}{\small (s)-t(s)}\\
{\small (5)} & {\small r}^{2}{\small x}_{0}{\small (r)+(1-r)}^{2}%
{\small x}_{1}{\small (r)-t(r)} & {\small \geq} & {\small r}^{2}{\small x}%
_{0}{\small (s)+(1-r)}^{2}{\small x}_{1}{\small (s)-t(s)}\\
{\small (6)} & {\small (1-r)}^{2}{\small x}_{0}{\small (1-r)+r}^{2}%
{\small x}_{1}{\small (1-r)-t(1-r)} & {\small \geq} & {\small (1-r)}%
^{2}{\small x}_{0}{\small (1-s)+r}^{2}{\small x}_{1}{\small (1-s)-t(1-s)}%
\end{array}
\]
Constraints (1)-(2) are the IR constraints of types $s$ and $1-s$. Constraints
(3)-(4) are the IC constraints that prevent them from mimicking one another.
Finally, constraints (5)-(6) are the IC constraints that prevent types $r$ and
$1-r$ from mimicking types $s$ and $1-s$, respectively.

We first establish that the optimal solution is symmetric in the sense that
for every $i,\beta$, $n_{1}=n_{0}$, $q_{i}(\beta)=q_{1-i}(1-\beta)$, and
$t(\beta)=t(1-\beta)$. To see why, assume the contrary. Note that if the
profile $(n_{i},q_{i}(\beta),t(\beta))_{i,\beta}$ is optimal (and therefore
satisfies all the constraints), then the permutation $n_{i}^{\prime}=n_{1-i}$,
$q_{i}^{\prime}(\beta)=q_{1-i}(1-\beta)$, and $t^{\prime}(\beta)=t(1-\beta)$,
is feasible and induces the same profit for the monopolist. Now define
$q_{i}^{\prime\prime}(\beta)$ such that
\[
x_{i}^{\prime\prime}(\beta)=\frac{1}{2}(x_{i}(\beta)+x_{i}^{\prime}(\beta))
\]
for every $i,\beta$. Since $x_{i}(\beta)$ is a one-to-one function of
$q_{i}(\beta)$, this defines $q^{\prime\prime}$ unambiguously. Likewise,
define%
\begin{align*}
t_{i}^{\prime\prime}(\beta)  &  =\frac{1}{2}(t_{i}(\beta)+t_{i}^{\prime}%
(\beta))\\
n_{i}^{\prime\prime}  &  =\frac{1}{2}(n_{i}+n_{i}^{\prime})
\end{align*}
Note that constraints (1)-(6) above are all linear in $x$ and $t$. Therefore,
since $(x,t)$ and $(x^{\prime},t^{\prime})$ satisfy the constraints, so does
$(x^{\prime\prime},t^{\prime\prime})$. Since $x_{i}(\beta)$ is strictly
concave in $q_{i}(\beta)$,%
\[
x_{i}^{\prime\prime}(\beta)=\frac{1}{2}\left[  \frac{q_{i}(\beta)}{q_{i}%
(\beta)+\sigma^{2}}+\frac{q_{1-i}(1-\beta)}{q_{1-i}(1-\beta)+\sigma^{2}%
}\right]  <\frac{\frac{1}{2}q_{i}(\beta)+\frac{1}{2}q_{1-i}(1-\beta)}{\frac
{1}{2}q_{i}(\beta)+\frac{1}{2}q_{1-i}(1-\beta)+\sigma^{2}}%
\]
Therefore,
\begin{align*}
q_{i}^{\prime\prime}(\beta)  &  <\frac{1}{2}q_{i}(\beta)+\frac{1}{2}%
q_{1-i}(1-\beta)\\
&  =\frac{1}{2}q_{i}(\beta)+\frac{1}{2}q_{i}^{\prime}(\beta)\\
&  \leq\frac{1}{2}n_{i}+\frac{1}{2}n_{i}^{\prime}\\
&  =n_{i}^{\prime\prime}%
\end{align*}
This means that $(q^{\prime\prime},t^{\prime\prime})$ replicates the
monopolist's revenue from $(q,t)$, but it can be attained with a strictly
lower total database than $n_{0}+n_{1}$ --- contradicting the optimality of
$(n,q,t)$ and $(n^{\prime},q^{\prime},t^{\prime})$. From now on, we use $N$ to
denote the value of $n_{0}=n_{1}$.

Now observe that by the symmetry, constraints (3)-(4) together imply
$x_{1}(s)\geq x_{1}(1-s)$ and (equivalently) $x_{0}(1-s)\geq x_{0}(s)$. Also
by symmetry, constraints (1), (3) and (5) are equivalent to constraints (2),
(4) and (6), respectively. Therefore, we can retain only the former. By
standard arguments, constraints (1) and (5) are binding, and type $r$'s data
access is efficient:
\[
x_{0}(r)=x_{1}(r)=\frac{N}{N+\sigma^{2}}%
\]
Plugging these in the objective function, and the constraint $x_{1}(s)\geq
x_{1}(1-s)$, we can restate the monopolist's problem as finding $N$,
$x_{1}(s)$, $x_{1}(1-s)$, to maximize%

\begin{align}
& \left[  \frac{1}{2}(1-s)^{2}-p_{r}(1-r)^{2}\right]  x_{1}(s)+\left[
\frac{1}{2}s^{2}-p_{r}r^{2}\right]  x_{1}(1-s)\label{reduced objective}\\
& +p_{r}[r^{2}+(1-r)^{2}]\frac{N}{N+\sigma^{2}}-2cN\nonumber
\end{align}
\bigskip subject to the constraint that $x_{1}(s)\geq x_{1}(1-s)$. This
constraint must bind in optimum: Since $s>r$ and $p(r)<\frac{1}{2}$,
expression (\ref{reduced objective}) strictly increases in $x_{1}(1-s)$. It
follows that $q_{1}(s)=q_{0}(s)=q_{1}(1-s)=q_{0}(1-s)=Q$.

The problem is thus to find $N$ and $Q\leq N$ to maximize%
\[
\left[  \frac{1}{2}(s^{2}+(1-s)^{2})-p(r)(r^{2}+(1-r)^{2})\right]  \frac
{Q}{Q+\sigma^{2}}+p(r)[r^{2}+(1-r)^{2}]\frac{N}{N+\sigma^{2}}-2cN
\]
Therefore, the solution for $Q$ is%
\[
Q=\left\{
\begin{array}
[c]{ccc}%
N & if & 2p(r)\leq\frac{s^{2}+(1-s)^{2}}{r^{2}+(1-r)^{2}}\\
0 & otherwise &
\end{array}
\right.
\]
This solution trivially satisfies the IR and IC constraints omitted from the
relaxed problem.

When $Q=N$, this means that all types get full access and pay an amount equal
to the \textquotedblleft low\textquotedblright\ types' willingness to pay:%
\[
t(\beta)=(s^{2}+(1-s)^{2})\frac{N}{N+\sigma^{2}}%
\]
for all $\beta$. When $Q=0$, this means that the \textquotedblleft
low\textquotedblright\ types $s$ and $1-s$ are entirely excluded (and
therefore pay nothing), while the \textquotedblleft high\textquotedblright%
\ types get full access and pay their willingness to pay:%
\[
t(r)=t(1-r)=(r^{2}+(1-r)^{2})\frac{N}{N+\sigma^{2}}%
\]
\ In either case, the optimal value of $N$ will be sub-optimally below the
first-best level, as can be easily verified by taking first-order conditions
in the first-best and the reduced second-best problems. $\blacksquare$%
\bigskip\bigskip

\noindent{\LARGE Appendix II: Other Derivations}\bigskip

\noindent{\large Derivation of Posterior Variances in Section 3}\bigskip

\noindent Recall the following independent Gaussian variables: $\mu\sim
N(0,\sigma_{\mu}^{2})$, $x_{t}\sim N(0,1)$ and $\varepsilon_{t,i}\sim
N(0,\sigma_{\varepsilon}^{2})$, where $t=0,1\ $and $i\in\{1,...,n_{t}\}.$ Also
recall that an observation $i$ from the period-$t$ sample is a realization
$y_{t,i}=\mu+x_{t}+\varepsilon_{t,i},$ and that types $S$ and $L$ are
interested in forecasting $\theta^{S}=\mu+x_{1}$ and $\theta^{L}=\mu$,
respectively. The prior variances over $\theta^{S}$ and $\theta^{L}$ are
$\sigma_{\mu}^{2}+1$ and $\sigma_{\mu}^{2}$, respectively.

From type $L$'s point of view, a period-$t$ sample generates a conditionally
independent signal $\bar{y}_{t}=\theta^{L}+x_{t}+\bar{\varepsilon}_{t}$, where
$\bar{\varepsilon}_{t}$ is the average observational noise in the period-$t$
sample. The variance of the period-$t$ signal conditional on $\theta^{L}$ is
$1+\sigma_{\varepsilon}^{2}/n_{t}$. From $S$'s point of view, the two periods'
samples generate the signals $\bar{y}_{1}=\theta^{S}+\bar{\varepsilon}_{1}$
and $\bar{y}_{0}=\theta^{S}+x_{0}-x_{1}+\bar{\varepsilon}_{0}$. We now
calculate the variance of the types' posterior beliefs.

For $c\in\{0,1\}$, we have the following joint normal distribution (where
$c=0$ gives us the joint distribution with $\mu$ as the first variable and
$c=1$ gives us the joint distribution with $\mu+x_{1}$ as the first
variable).
\[%
\begin{pmatrix}
\mu+cx_{1}\\
\bar{y}_{0}\\
\bar{y}_{1}%
\end{pmatrix}
\sim N\left(
\begin{pmatrix}
0\\
0\\
0
\end{pmatrix}
,%
\begin{pmatrix}
\sigma_{\mu}^{2}+c & \sigma_{\mu}^{2} & \sigma_{\mu}^{2}+c\\
\sigma_{\mu}^{2} & \sigma_{\mu}^{2}+1+\frac{\sigma_{\varepsilon}^{2}}{n_{0}} &
\sigma_{\mu}^{2}\\
\sigma_{\mu}^{2}+c & \sigma_{\mu}^{2} & \sigma_{\mu}^{2}+1+\frac
{\sigma_{\varepsilon}^{2}}{n_{1}}%
\end{pmatrix}
\right)  .
\]
Denote
\[
A=%
\begin{pmatrix}
\sigma_{\mu}^{2}+1+\frac{\sigma_{\varepsilon}^{2}}{n_{0}} & \sigma_{\mu}^{2}\\
\sigma_{\mu}^{2} & \sigma_{\mu}^{2}+1+\frac{\sigma_{\varepsilon}^{2}}{n_{1}}%
\end{pmatrix}
\]
Then,%

\begin{equation}
det(A)=(\sigma_{\mu}^{2}+1+\frac{\sigma_{\varepsilon}^{2}}{n_{0}})(\sigma
_{\mu}^{2}+1+\frac{\sigma_{\varepsilon}^{2}}{n_{1}})-\sigma_{\mu}^{4}
\label{det_A}%
\end{equation}
and%

\begin{equation}
A^{-1}=%
\begin{pmatrix}
\sigma_{\mu}^{2}+1+\frac{\sigma_{\varepsilon}^{2}}{n_{1}} & -\sigma_{\mu}%
^{2}\\
-\sigma_{\mu}^{2} & \sigma_{\mu}^{2}+1+\frac{\sigma_{\varepsilon}^{2}}{n_{0}}%
\end{pmatrix}
(det(A))^{-1}.\label{A_inverse}%
\end{equation}
Therefore,%
\[
Var(\mu+cx_{1}\mid\bar{y}_{0},\bar{y}_{1})=\sigma_{\mu}^{2}+c-%
\begin{pmatrix}
\sigma_{\mu}^{2} & \sigma_{\mu}^{2}+c
\end{pmatrix}
A^{-1}%
\begin{pmatrix}
\sigma_{\mu}^{2}\\
\sigma_{\mu}^{2}+c
\end{pmatrix}
\]
Plugging (\ref{A_inverse}) and $\sigma_{\varepsilon}^{2}=1$ into this
expression, and then simplifying, yields the desired expressions for each
type's variance reduction. For $c=0$, we have%
\[
Var(\mu)-Var(\mu\mid\bar{y}_{0},\bar{y}_{1})=\frac{\sigma_{\mu}^{4}%
(n_{1}+n_{0}+2n_{0}n_{1})}{\sigma_{\mu}^{2}(n_{1}+n_{0}+2n_{0}n_{1}%
)+(1+n_{0})(1+n_{1})}.
\]
For $c=1$, we have%
\[
Var(\mu+x_{1})-Var(\mu+x_{1}\mid\bar{y}_{0},\bar{y}_{1})=\frac{\sigma_{\mu
}^{4}(n_{1}+n_{0}+2n_{0}n_{1})+3\sigma_{\mu}^{2}n_{0}n_{1}+2\sigma_{\mu}%
^{2}n_{1}+n_{1}+n_{0}n_{1}}{\sigma_{\mu}^{2}(n_{1}+n_{0}+2n_{0}n_{1}%
)+(1+n_{0})(1+n_{1})}.
\]

\pagebreak

\noindent{\large Proof of Remark \textbf{\ref{remark properties}}}\bigskip

\noindent\textbf{Proof of (i).} This follows from noting that%
\begin{align}
\frac{\partial}{\partial n_{0}}V_{L}(n_{0},n_{1})  &  =\frac{\sigma_{\mu}%
^{4}\left(  n_{1}+1\right)  ^{2}}{\left(  n_{0}+n_{1}+n_{0}n_{1}+\sigma_{\mu
}^{2}n_{0}+\sigma_{\mu}^{2}n_{1}+2\sigma_{\mu}^{2}n_{0}n_{1}+1\right)  ^{2}%
}>0\label{dVL_dn0}\\
\frac{\partial}{\partial n_{1}}V_{L}(n_{0},n_{1})  &  =\frac{\sigma_{\mu}%
^{4}\left(  n_{0}+1\right)  ^{2}}{\left(  n_{0}+n_{1}+n_{0}n_{1}+\sigma_{\mu
}^{2}n_{0}+\sigma_{\mu}^{2}n_{1}+2\sigma_{\mu}^{2}n_{0}n_{1}+1\right)  ^{2}}>0
\label{dVL_dn1}%
\end{align}

\begin{align}
\frac{\partial}{\partial n_{0}}V_{S}(n_{0},n_{1})  &  =\frac{\sigma_{\mu}^{4}%
}{\left(  n_{0}+n_{1}+n_{0}n_{1}+\sigma_{\mu}^{2}n_{0}+\sigma_{\mu}^{2}%
n_{1}+2\sigma_{\mu}^{2}n_{0}n_{1}+1\right)  ^{2}}>0\label{dVSdn0}\\
\frac{\partial}{\partial n_{1}}V_{S}(n_{0},n_{1})  &  =\frac{\left(
n_{0}+\sigma_{\mu}^{2}+2\sigma_{\mu}^{2}n_{0}+1\right)  ^{2}}{\left(
n_{0}+n_{1}+n_{0}n_{1}+\sigma_{\mu}^{2}n_{0}+\sigma_{\mu}^{2}n_{1}%
+2\sigma_{\mu}^{2}n_{0}n_{1}+1\right)  ^{2}}>0 \label{dVSdn1}%
\end{align}

\noindent\textbf{Proof of (ii).} We begin by verifying that $V_{L}(n_{0}%
,n_{1})$ is strictly concave. Its Hessian matrix is given by%
\[%
\begin{array}
[c]{cc}%
\frac{\partial^{2}}{\partial\left(  n_{0}\right)  ^{2}}V_{L}(n_{0},n_{1}) &
\frac{\partial^{2}}{\partial n_{1}\partial n_{0}}V_{L}(n_{0},n_{1})\\
\frac{\partial^{2}}{\partial n_{1}\partial n_{0}}V_{L}(n_{0},n_{1}) &
\frac{\partial^{2}}{\partial\left(  n_{1}\right)  ^{2}}V_{L}(n_{0},n_{1})
\end{array}
\]
The expressions for the terms in each cell are as follows:%
\begin{align*}
\frac{\partial^{2}}{\partial\left(  n_{0}\right)  ^{2}}V_{L}(n_{0},n_{1})  &
=\frac{-2\sigma_{\mu}^{4}\left(  n_{1}+1\right)  ^{2}\left(  n_{1}+\sigma
_{\mu}^{2}+2\sigma_{\mu}^{2}n_{1}+1\right)  }{\left(  n_{0}+n_{1}+n_{0}%
n_{1}+\sigma_{\mu}^{2}n_{0}+\sigma_{\mu}^{2}n_{1}+2\sigma_{\mu}^{2}n_{0}%
n_{1}+1\right)  ^{3}}\\
\frac{\partial}{\partial n_{0}\partial n_{1}}V_{L}(n_{0},n_{1})  &
=\frac{-2\sigma_{\mu}^{6}\left(  n_{0}+1\right)  \left(  n_{1}+1\right)
}{\left(  n_{0}+n_{1}+n_{0}n_{1}+\sigma_{\mu}^{2}n_{0}+\sigma_{\mu}^{2}%
n_{1}+2\sigma_{\mu}^{2}n_{0}n_{1}+1\right)  ^{3}}\\
\frac{\partial^{2}}{\partial\left(  n_{1}\right)  ^{2}}V_{L}(n_{0},n_{1})  &
=\frac{-2\sigma_{\mu}^{4}\left(  n_{0}+1\right)  ^{2}\left(  n_{0}+\sigma
_{\mu}^{2}+2\sigma_{\mu}^{2}n_{0}+1\right)  }{\left(  n_{0}+n_{1}+n_{0}%
n_{1}+\sigma_{\mu}^{2}n_{0}+\sigma_{\mu}^{2}n_{1}+2\sigma_{\mu}^{2}n_{0}%
n_{1}+1\right)  ^{3}}%
\end{align*}
$\allowbreak$The function $V_{L}(n_{0},n_{1})$ is strictly concave if its
Hessian matrix is negative definite. To confirm this, note first that the
first principal minor is negative: $\frac{\partial^{2}}{\partial\left(
n_{0}\right)  ^{2}}V_{L}(n_{0},n_{1})<0.$ Second, note that the determinant of
the Hessian matrix is positive:%

\begin{align*}
&  \frac{\partial^{2}}{\partial\left(  n_{0}\right)  ^{2}}V_{L}(n_{0}%
,n_{1})\cdot\frac{\partial^{2}}{\partial\left(  n_{1}\right)  ^{2}}V_{L}%
(n_{0},n_{1})-\left(  \frac{\partial^{2}}{\partial n_{1}\partial n_{0}}%
V_{L}(n_{0},n_{1})\right)  ^{2}\\
&  =\frac{4\sigma_{\mu}^{8}\left(  n_{0}+1\right)  ^{2}\left(  n_{1}+1\right)
^{2}\left(  \left(  n_{1}+\sigma_{\mu}^{2}+2\sigma_{\mu}^{2}n_{1}+1\right)
\left(  n_{0}+\sigma_{\mu}^{2}+2\sigma_{\mu}^{2}n_{0}+1\right)  -\sigma_{\mu
}^{4}\right)  }{\left(  n_{0}+n_{1}+n_{0}n_{1}+\sigma_{\mu}^{2}n_{0}%
+\sigma_{\mu}^{2}n_{1}+2\sigma_{\mu}^{2}n_{0}n_{1}+1\right)  ^{6}}\\
&  >0
\end{align*}

We next turn to verifying that $V_{S}(n_{0},n_{1})$ is strictly concave. Its
Hessian matrix is%
\[%
\begin{array}
[c]{cc}%
\frac{\partial^{2}}{\partial\left(  n_{0}\right)  ^{2}}V_{S}(n_{0},n_{1}) &
\frac{\partial^{2}}{\partial n_{1}\partial n_{0}}V_{S}(n_{0},n_{1})\\
\frac{\partial^{2}}{\partial n_{1}\partial n_{0}}V_{S}(n_{0},n_{1}) &
\frac{\partial^{2}}{\partial\left(  n_{1}\right)  ^{2}}V_{S}(n_{0},n_{1})
\end{array}
\]
The expressions for the terms in each cell are as follows:%

\begin{align*}
\frac{\partial^{2}}{\partial\left(  n_{0}\right)  ^{2}}V_{S}(n_{0},n_{1})  &
=\frac{-2\sigma_{\mu}^{4}\left(  n_{1}+\sigma_{\mu}^{2}+2\sigma_{\mu}^{2}%
n_{1}+1\right)  }{\left(  n_{0}+n_{1}+n_{0}n_{1}+\sigma_{\mu}^{2}n_{0}%
+\sigma_{\mu}^{2}n_{1}+2\sigma_{\mu}^{2}n_{0}n_{1}+1\right)  ^{3}}\\
\frac{\partial}{\partial n_{0}\partial n_{1}}V_{S}(n_{0},n_{1})  &
=\frac{-2\sigma_{\mu}^{4}\left(  n_{0}+\sigma_{\mu}^{2}+2\sigma_{\mu}^{2}%
n_{0}+1\right)  }{\left(  n_{0}+n_{1}+n_{0}n_{1}+\sigma_{\mu}^{2}n_{0}%
+\sigma_{\mu}^{2}n_{1}+2\sigma_{\mu}^{2}n_{0}n_{1}+1\right)  ^{3}}\\
\frac{\partial^{2}}{\partial\left(  n_{1}\right)  ^{2}}V_{S}(n_{0},n_{1})  &
=\frac{-2\left(  n_{0}+\sigma_{\mu}^{2}+2\sigma_{\mu}^{2}n_{0}+1\right)  ^{3}%
}{\left(  n_{0}+n_{1}+n_{0}n_{1}+\sigma_{\mu}^{2}n_{0}+\sigma_{\mu}^{2}%
n_{1}+2\sigma_{\mu}^{2}n_{0}n_{1}+1\right)  ^{3}}%
\end{align*}
$V_{S}(n_{0},n_{1})$ is strictly concave since $\frac{\partial^{2}}%
{\partial\left(  n_{0}\right)  ^{2}}V_{S}(n_{0},n_{1})<0$ --- i.e., the first
principal minor is negative --- and the determinant of the Hessian matrix is positive:%

\begin{align*}
&  \frac{\partial^{2}}{\partial\left(  n_{0}\right)  ^{2}}V_{S}(n_{0}%
,n_{1})\cdot\frac{\partial^{2}}{\partial\left(  n_{1}\right)  ^{2}}V_{S}%
(n_{0},n_{1})-\left(  \frac{\partial^{2}}{\partial n_{1}\partial n_{0}}%
V_{S}(n_{0},n_{1})\right)  ^{2}\\
&  =\frac{4\sigma_{\mu}^{4}\left(  n_{0}+\sigma_{\mu}^{2}+2\sigma_{\mu}%
^{2}n_{0}+1\right)  ^{2}\left[  \left(  n_{1}+\sigma_{\mu}^{2}+2\sigma_{\mu
}^{2}n_{1}+1\right)  \left(  n_{0}+\sigma_{\mu}^{2}+2\sigma_{\mu}^{2}%
n_{0}+1\right)  -\sigma_{\mu}^{4}\right]  }{\left(  n_{0}+n_{1}+n_{0}%
n_{1}+\sigma_{\mu}^{2}n_{0}+\sigma_{\mu}^{2}n_{1}+2\sigma_{\mu}^{2}n_{0}%
n_{1}+1\right)  ^{6}}\\
&  >0
\end{align*}

\noindent\textbf{Proof of (iii).} From inspection of (\ref{VL}) it is easy to
see that $V_{L}(x,y)=V_{L}(y,x).$ To see that $V_{S}(x,y)>V_{S}(y,x)$ for
$y>x$, note that%
\[
V_{S}(x,y)-V_{S}(y,x)=\frac{\left(  y-x\right)  \left(  2\sigma_{\mu}%
^{2}+1\right)  }{\sigma_{\mu}^{2}(y+x+2xy)+(1+x)(1+y)}>0
\]
The observation that $\frac{\partial V_{S}(n_{0},n_{1})}{\partial n_{1}}%
>\frac{\partial V_{S}(n_{0},n_{1})}{\partial n_{0}}$ follows from comparing
equation (\ref{dVSdn0}) to equation (\ref{dVSdn1}).

\noindent\textbf{Proof of (iv).} Follows immediately from equations (\ref{VL})
and (\ref{VS}).

\noindent\textbf{Proof of (v).} Follows immediately from equations
(\ref{dVL_dn0})-(\ref{dVSdn1}).

\end{document}